\documentclass[prb,aps,showpacs,twocolumn,reprint,superscriptaddress,english]{revtex4}
\usepackage{graphics,bm}
\usepackage{amssymb,ulem}
\usepackage{epsfig}
\usepackage[usenames]{color}
\usepackage{float}
\usepackage{epstopdf}
\usepackage[T1]{fontenc}
\usepackage[latin9]{inputenc}
\usepackage{amsmath}
\usepackage{babel}

\begin{document}

\title{Green's function formalism for spin transport in metal-insulator-metal heterostructures}

\author{Jiansen Zheng}
\affiliation{Institute for Theoretical Physics and Center for Extreme Matter and Emergent
	Phenomena, Utrecht University, Leuvenlaan 4, 3584 CE Utrecht, The Netherlands}
\author{Scott Bender}
\affiliation{Institute for Theoretical Physics and Center for Extreme Matter and Emergent
	Phenomena, Utrecht University, Leuvenlaan 4, 3584 CE Utrecht, The Netherlands}
\author{Jogundas Armaitis}
\affiliation{Institute of Theoretical Physics and Astronomy, Vilnius University,
	Saul\.etekio Ave.~3, LT-10222 Vilnius, Lithuania}
\author{Roberto E. Troncoso}
\affiliation{Department of Physics, Norwegian University of Science and Technology, NO-7491 Trondheim, Norway}
\affiliation{Departamento de F\'isica, Universidad T\'ecnica Federico Santa Mar\'ia, Avenida Espa\~na 1680, Valpara\'iso, Chile}
\author{Rembert A. Duine}
\affiliation{Institute for Theoretical Physics and Center for Extreme Matter and Emergent
	Phenomena, Utrecht University, Leuvenlaan 4, 3584 CE Utrecht, The Netherlands}
\affiliation{Department of Applied Physics, Eindhoven University of Technology, PO Box 513,
	5600 MB Eindhoven, The Netherlands}
\date{\today}

\begin{abstract}
We develop a Green's function formalism for spin transport through heterostructures that contain metallic leads and insulating ferromagnets. While this formalism in principle allows for the inclusion of various magnonic interactions, we focus on Gilbert damping. As an application, we consider ballistic spin transport by exchange magnons in a metal-insulator-metal heterostructure with and without disorder. For the former case, we show that the interplay between disorder and Gilbert damping leads to spin current fluctuations. For the case without disorder, we obtain the dependence of the transmitted spin current on the thickness of the ferromagnet. Moreover, we show that the results of the Green's function formalism agree in the clean and continuum limit with those obtained from the linearized stochastic Landau-Lifshitz-Gilbert equation. The developed Green's function formalism is a natural starting point for numerical studies of magnon transport in heterostructures that contain normal metals and magnetic insulators. 
\end{abstract}


\maketitle

\def\bX{{\bm X}}
\def\bx{{\bm x}}
\def\bk{{\bm k}}
\def\bK{{\bm K}}
\def\bq{\mathbf{q}}
\def\br{{\bm r}}
\def\bp{{\bm p}}
\def\half{\frac{1}{2}}
\def\args{(\bm, t)}

\section{Introduction}
Magnons are the bosonic quanta of spin waves, oscillations in the magnetization orientation in magnets \cite{landau1958statistical,kittel1966introduction}. Interest in magnons has recently revived as enhanced experimental control has made them attractive as potential data carriers of spin information over long distances and without Ohmic dissipation  \cite{chumak2015magnon}.
In general, magnons exist in two regimes. One is the dipolar magnon with long wavelengths that is dominated by long-range dipolar interactions and which can be generated e.g. by ferromagnetic resonance \cite{kasuya1961relaxation,sparks1970ferromagnetic}. The other type is the exchange magnon \cite{sandweg2011spin}, dominated by exchange interactions and which generally has higher frequency and therefore perhaps more potential for applications in magnon based devices \cite{chumak2015magnon}. In this paper, we focus on transport of exchange magnons.
 
Thermally driven magnon transport has been widely investigated, and is closely related to spin pumping of spin currents across the interface between insulating ferromagnets (FMs) and normal metals (NM) \cite{vlietstra2016detection,talalaevskij2017magnetic,holanda2017simultaneous} and detection of spin current by the inverse spin Hall Effect  \cite{saitoh2006conversion}. The most-often studied thermal effect in this context is the spin Seebeck effect, which is the generation of a spin current by a temperature gradient applied to a magnetic insulator that is detected in an adjacent normal metal via the inverse spin Hall effect
\cite{uchida2008observation,xiao2010theory}.  Here, thermal fluctuations in the NM contacts drive spin transport into the FM, while the dissipation of spin back into the NM by magnetic dynamics is facilitated by the above mentioned spin-pumping mechanism.

 The injection of spin into a FM can also be accomplished electrically, via the interaction of spin polarized electrons in the NM and the localized magnetic moments of the FM.  Reciprocal to spin-pumping is the spin-transfer torque, which, in the presence of a spin accumulation (typically generated by the spin Hall effect) in the NM, drives magnetic dynamics in the FM\cite{berger1996emission,slonczewski1996current}. Spin pumping likewise underlies the flow of spin back into the NM contacts, which serve as magnon reservoirs. In two-terminal set-ups based on YIG and Pt, the characteristic length scales and device-specific parameter dependence of magnon transport has attracted enormous attention, both in experiments and theory.  Cornelissen {\it et al.} \cite{cornelissen2015long}  studied the excitation and detection of high-frequency magnons in YIG and measured the propagating length of magnons, which reaches up to $10$ $\mu$m in a thin YIG film at room temperature. Other experiments have shown that the polarity reversal of detected spins of thermal magnons in non-local devices of YIG are strongly dependent on temperature, YIG film thickness, and injector-detector
separation distance \cite{zhou2017lateral}.  That the interfaces are crucial can e.g. be seen by changing the interface electron-magnon coupling, which was found to 
significantly alter the longitudinal spin Seebeck effect \cite{guo2016influence}.

A linear-response transport theory was developed for diffusive spin and heat transport by magnons in magnetic insulators with metallic contacts. Among other quantities, this theory is parameterized by relaxation lengths for the magnon chemical potential and magnon-phonon energy relaxation \cite{cornelissen2016magnon,flebus2016two}. In a different but closely-related development, Onsager relations for the magnon spin and heat currents driven by magnetic field and temperature differences were established for insulating ferromagnet junctions, and a magnon analogue of the Wiedemann-Franz law was is also predicted \cite{nakata2015wiedemann,nakata2017spin}. Wang {\it et al.} \cite{wang2004spin} consider ballistic transport of magnons through magnetic insulators with magnonic reservoirs --- rather than the more experimentally relevant situation of metallic reservoirs considered here --- and use a nonequilibrium Green's function formalism (NEGF) to arrive at Landuaer-B\"utikker-type expressions for the magnon current. The above-mentioned works are either in the linear-response regime or do not consider Gilbert damping and/or metallic reservoirs.  So far, a complete quantum mechanical framework to study exchange magnon transport through heterostructures containing metallic reservoirs that can access different regimes, ranging from ballistic to diffusive, large or small Gilbert damping, and/or small or large interfacial magnon-electron coupling, and that can incorporate Gilbert damping, is lacking. 

\begin{figure}[h!]
	\begin{center}
		\includegraphics[scale=0.25]{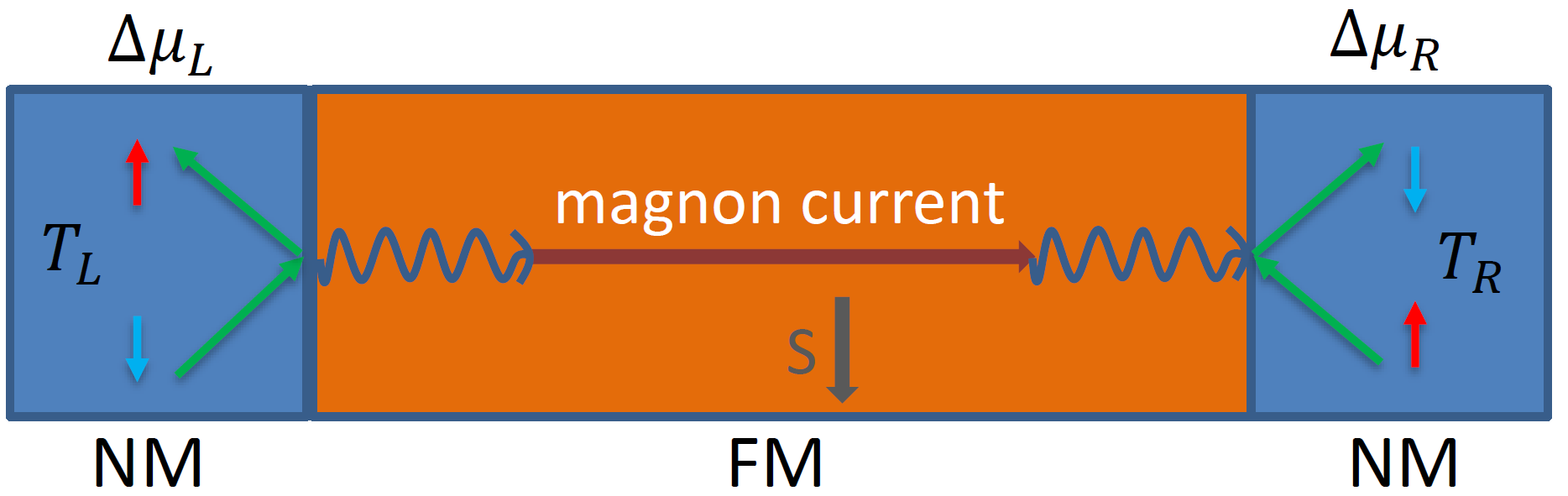}
		\caption{Illustration of the system where magnon transport in a ferromagnet (orange region) is driven by a spin accumulation difference $ \Delta \mu_L- \Delta \mu_R$ and temperature difference $T_L-T_R$ between two normal-metal leads (blue regions). Spin-flip scattering at the interface converts electronic to magnonic spin current. Here, $\bf S$ is the local spin density in equilibrium.}
		\label{fig:device}
	\end{center}
\end{figure}

In this paper we develop the non-equilibrium Green's function formalism \cite{di2008electrical} for spin transport through NM-FM-NM heterostructures (see Fig.~\ref{fig:device}). In principle, this formalism straightforwardly allows for adding arbitrary interactions, such as scattering of magnons with impurities and phonons, Gilbert damping, and magnon-magnon interactions, and provides a suitable platform to study magnon spin transport numerically, in particular beyond linear response. Here, we apply the formalism to ballistic magnon transport through a low-dimensional channel in the presence of Gilbert damping. For that case, we compute the magnon spin current as a function of channel length both numerically and analytically. For the clean case in the continuum limit we show how to recover our results from the linearized stochastic Landau-Lifshitz-Gilbert (LLG) equation \cite{brown1963thermal} used previously to study thermal magnon transport in the ballistic regime  \cite{hoffman2013landau} that applies to to clean systems at low temperatures such that Gilbert damping is the only relaxation mechanism. Using this formalism we also consider the interplay between Gilbert damping and disorder and show that it leads to spin-current fluctuations. 

This paper is organized as follows. 
In Sec.~\ref{negf}, we discuss the non-equilibrium Green's function approach to magnon transport and derive an expression for the magnon spin current. Additionally a Landauer-B\"uttiker formula for the magnon spin current is derived.
In Sec.~\ref{numerical}, we illustrate the formalism by numerically considering ballistic magnon transport and determine the dependence of the spin current on thickness of the ferromagnet. To further understand these numerical results, we consider the formalism analytically in the continuum limit in Sec.~\ref{sec:analytical}, and also show that in that limit we obtain the same results using the stochastic LLG equation. We give a further discussion and outlook in section \ref{discussion}.

\section{Non-equilibrium Green's function formalism}
\label{negf}
In this section we describe our model and, using Keldysh theory, arrive at an expression for the density matrix of the magnons from which all observables can be calculated. The reader interested in applying the final result of our formalism may skip ahead to Sec.~\ref{sec:summary} where we give a summary on how to implement it. 
\subsection{Model}
\begin{figure}[!h]
	\includegraphics[scale=1.15,trim=20 4 4 4,clip]{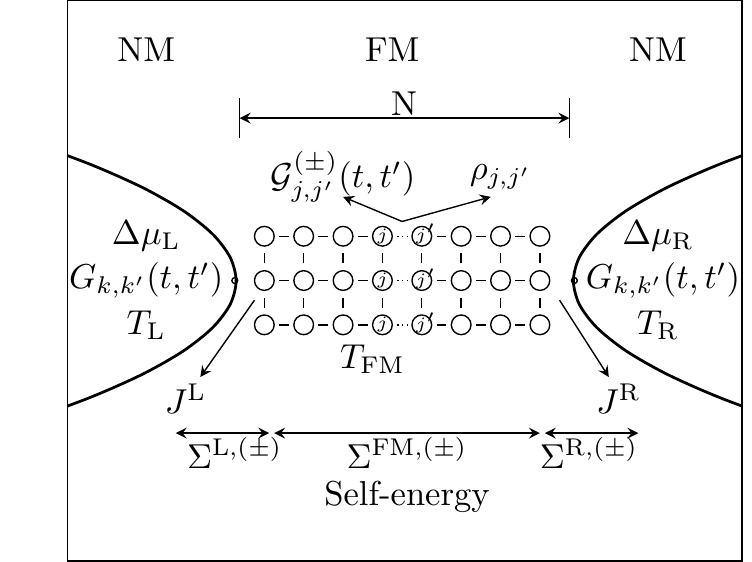} 
	\caption{Schematic for the NM-FM-NM heterostructure and notation for the Green's functions and self-energies. The array of circles denotes the localized magnetic moments, while the two regions outside the parabolic lines denote the leads, i.e., reservoirs of polarized electrons.
		Moreover, $J_{j;kk'}^{L/R}$ denotes the interface coupling, and $T_{L/R}$ and $
		 \Delta \mu_{L/R}$ denote the temperature and spin accumulation for the leads. The properties of the magnons are encoded in $ {\mathcal G}_{j,j'}^{(+)}(t,t')$, the retarded magnon Green's function, and the magnon density matrix $\rho_{j,j'}$. The number of sites in the spin-current direction is $N$. The self-energies $\Sigma^{FM,(\pm)}$, $\Sigma^{L,(\pm)}$, $\Sigma^{R,(\pm)}$ are due to Gilbert damping, and the left and right lead, respectively.
	}
	\label{fig:schematic}
\end{figure}

We consider a magnetic insulator connected to two nonmagnetic metallic leads, as shown in Fig.~\ref{fig:schematic}. For our formalism it is most convenient to consider both the magnons and the electrons as hopping on the lattice for the ferromagnet. Here, we consider the simplest versions of such cubic lattice models; extensions, e.g. to multiple magnon and/or electron bands, and multiple leads are straightforward. The leads have a temperature $T_{L/R}$ and a spin accumulation $\Delta \mu_{L/R}$ that injects spin current from the non-magnetic metal into the magnetic insulator. This nonzero spin accumulation could, e.g., be established by the spin Hall effect. 

The total Hamiltonian is a sum of the uncoupled magnon and lead Hamiltonians together with a coupling term:
\begin{equation}
\hat{H}_{\rm tot}= \hat{H}_{\rm FM}  + \hat H_{\rm NM} + \hat H_{C}~.
\end{equation}
Here, $\hat{H}_{\rm FM}$ denotes the free Hamiltonian for the magnons,
\begin{equation}
\label{eq:magnonham}
\hat{H}_{\rm FM} = -\sum_{<j,j'>} J_{j,j'} b_{j'}^{\dagger} b_{j}+\sum_j \Delta_j b_{j}^{\dagger} b_{j} \equiv \sum_{<j,j'>} h_{j',j} b_{j'}^{\dagger} b_{j}~.
\end{equation}
where $ b_j (b^{\dagger}_{j})$ is a magnon annihilation (creation) operator.  This hamiltonian can be derived from a spin hamiltonian using the Holstein-Primakoff transformation  \cite{holstein1940field, auerbach2012interacting} and expanding up to second order in the bosonic fields. Eq.~(\ref{eq:magnonham}) describes hopping of the magnons with amplitude $J_{j,j'}$ between sites labeled by $j$ and $j'$ on the lattice, with an on-site potential energy $\Delta_j$ that, if taken to be homogeneous, would correspond to the magnon gap induced by a magnetic field and anisotropy. We have taken the external field in the $-z$ direction, so that one magnon, created at site $j$ by the operator $\hat b^\dagger_j$, corresponds to spin $+\hbar$. 

The Hamiltonian for the electrons in the leads is
\begin{equation}
\hat H_{\rm NM} = - \sum_{r \in \{L,R \}}\sum_{<k,k'>}\sum_{ \sigma \in {\uparrow,\downarrow}} t^r \hat \psi^\dagger_{k\sigma r}  \hat \psi_{k'\sigma r}+h.c. 
\end{equation}
where the electron creation ($\psi^\dagger_{k \sigma r}$) and annihilation ($\psi_{k \sigma r}$) operators are labelled by the lattice position $k$, spin $\sigma$, and an index $r$ distinguishing (L)eft and (R)ight leads. The hopping amplitude for the electrons is denoted by $t^r$ and could in principle be different for different leads. Moreover, terms to describe hopping beyond nearest neighbor can be straightforwardly included. Below we will show that 
microscopic details will be incorporated in a single parameter per lead that describes the coupling between electrons and magnons. 

Finally, the Hamiltonian that describes the coupling between metal and insulator, $\hat H_{C}$, is given by
\cite{bender2012electronic}
\begin{equation}
\label{eq:interfacecouplhamiltonian}
\hat H_C = \sum_{r,j;k k'} \left( J^r_{j;k k'} \hat b^\dagger_j \hat \psi^\dagger_{k\downarrow r}  \hat \psi_{k'\uparrow r} + {\rm h.c.}\right)~,
\end{equation}
with the matrix elements $J^r_{j;k k'}$ that depend on the microscopic details of the interface.
An electron spin that flips from up to down at the interface creates one magnon with spin $+\hbar $ in the magnetic insulator. This form of coupling between electrons and magnons  derives from interface exchange coupling between spins in the insulators with electronic spins in the metal \cite{bender2012electronic}.

\begin{figure}[htb]
	\includegraphics[scale=1.5]{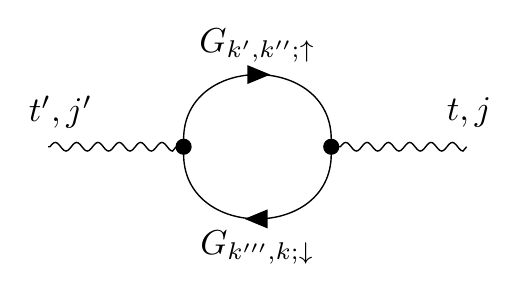}\label{fig:MagfeynD} 
	\caption{Feynman diagram for the spin-flip processes emitting and absorbing magnons that are represented by the wavy lines.  The two vertices indicate the exchange coupling at one of the interfaces of the magnetic insulator (sites $j,j'$) and normal metal (sites $k,k',k'',k'''$). $G_{k' k'';\uparrow}$ and  $G_{k''' k;\downarrow}$ denotes the electron Keldysh Green's function of one of the leads. }
	\label{fig:diagram}
\end{figure}

\subsection{Magnon density matrix and current}
\label{DensityMatrix}

Our objective is to calculate the steady-state magnon Green's function $i{\mathcal G}^{<}_{j,j'} (t,t') = \langle \hat b^\dagger_{j'} (t') \hat b_j (t)\rangle$, from which all observables are calculated (note that time-dependent operators refer to the Heisenberg picture). This ``lesser'' Green's function follows from the Keldysh Green's function
\begin{equation}
\label{eq:defgfkeldysh}
i {\mathcal G}_{j,j'}(t,t') \equiv {\rm Tr}
\left[ \hat \rho (t_0) T_{\mathcal C^{\infty}} \left( \hat b_{j} (t) \hat b^\dagger_{j'} (t')
\right)\right]~,
\end{equation}
with $\hat \rho (t_0)$ the initial (at time $t_0$) density matrix, and $\mathcal C^{\infty}$ the Keldysh contour, and ${\rm Tr}[...]$ stands for performing a trace average. The time-ordering operator on this contour is defined by
\begin{equation}
T_{\mathcal C^{\infty}} \left( \hat O(t) \hat O' (t')
\right) \equiv \theta (t,t') \hat O (t) \hat O' (t')
\pm \theta (t',t) \hat O' (t') \hat O
(t)~,
\end{equation}
with $\theta (t,t')$ the corresponding Heaviside step function 
and the $+ (-)$ sign applies when the operators have bosonic (fermionic)
commutation relations. In Fig.~\ref{fig:schematic} 
we schematically indicate the relevant quantities entering our theory.

At $t = 0$, the spin accumulation in the two leads is 

We compute the magnon self energy due the coupling between magnons and electrons to second order in the coupling matrix elements $J_{j;kk'}$. This implies that the magnons acquire a Keldysh self-energy due to lead $r$ given by
\begin{eqnarray}
\label{eq:keldyshsigmaleads}
\hbar \Sigma^r_{j,j'} (t,t') &=& -\frac{i}{\hbar} \sum_{k k' k'' k'''} J^r_{j;k k'} \left(J^r\right)^{*}_{j';k'' k'''} \nonumber \\
&& \times G_{k' k'';r\uparrow} (t,t') G_{k''' k;r \downarrow }(t',t)~,
\end{eqnarray}
where $ G_{k' k'';r\sigma} (t,t')$ denotes the electron Keldysh Green's function of lead $r$, that reads 
\begin{eqnarray}
\label{eq:electronkeldyshgreensfct}
G_{k k^{\prime};r\sigma}(t,t^{\prime})  & =& -i \langle T_{\mathcal C^{\infty}} \hat{\psi}_{k r \sigma}(t) \hat{\psi}^{\dagger}_{k^{\prime} r \sigma}(t^{\prime}) \rangle~.
\end{eqnarray} 
The Feynman diagram for this self-energy is shown in Fig.~\ref{fig:diagram}. While this self-energy is computed to second order in $J^r_{j;kk'}$, the magnon Green's function and the magnon spin current, both of which we evaluate below, contain all orders in $J^r_{j;kk'}$, which therefore does not need to be small. In this respect, our approach is different from the work of Ohnuma {\it et al.} [\onlinecite{ohnuma2017}], who evaluate the interfacial spin current to second order in the electron-magnon coupling. Irreducible diagrams other than that in Fig.~\ref{fig:diagram} involve one or more magnon propagators as internal lines and therefore correspond to magnon-magnon interactions at the interface induced by electrons in the normal metal. For the small magnon densities of interest to use here these can be safely neglected and the self-energy in Eq.~(\ref{eq:keldyshsigmaleads}) thus takes into account the dominant process of spin transfer between metal and insulator. 
 
The lesser and greater component of the electronic Green's functions can be expressed in terms of the spectral functions $A_{k k';r} (\epsilon)$ via
\begin{eqnarray}
- i G^{<}_{k k';r\sigma} &=& A_{k k';r} (\epsilon) N_F \left(\frac{\epsilon-\mu_{r\sigma}}{k_B T_r}\right)~; \nonumber \\
i G^{>}_{k k';r\sigma} &=& A_{k k';r} (\epsilon) \left[ 1-N_F \left(\frac{\epsilon-\mu_{r\sigma}}{k_B T_r}\right)\right]~,
\end{eqnarray}
with $N_F (x) = [e^{ x}+1]^{-1}$ the Fermi distribution function, $T_r$ the temperature of lead $r$ ($k_B$ being Boltzmann's constant) and $\mu_{\sigma,r}$ the chemical potential of spin projection $\sigma$ in lead $r$. As we will see later on, the lead chemical potential are taken spin-dependent to be able to implement nonzero spin accumulation. The spectral function is related to the retarded Green's function via
\begin{equation}
A_{k k';r} (\epsilon) = -2 {\rm Im} \left[ G^{(+)}_{k k';r} (\epsilon)\right]~,
\end{equation}
which does not depend on spin as the leads are taken to be normal metals. While the retarded Green's function of the leads can be determined explicitly for the model that we consider here, we will show below that such a level of detail is not needed but that, instead, we can parameterize the electron-magnon coupling by an effective interface parameter. 

As mentioned before, all steady-state properties of the magnon system are determined by the magnon lesser Green's function leading to the magnon density matrix. It is ultimately given by the kinetic equation \cite{di2008electrical,maciejko2007introduction}
\begin{equation}
\rho_{j,j'} \equiv \langle \hat b^\dagger_{j'} (t) \hat b_j (t)\rangle = \int \frac{d\epsilon}{(2\pi)}
\left[ {\mathcal G}^{(+)} (\epsilon) i \hbar \Sigma^{<} (\epsilon) {\mathcal G}^{(-)} (\epsilon)\right]_{j,j'}~,
\label{eq:dm_magnon}
\end{equation}
where $\hbar \Sigma_{j,j'} (t,t')$ is the total magnon self-energy  discussed in detail below, of which the "lesser" component enters in the above equation.  
In the above and what follows, quantities with suppressed site indexes are interpreted as matrices, and matrix multiplication applies for products of these quantities. The retarded $(+)$ and advanced $(-)$ magnon Green's functions satisfy
\begin{equation}
\left[ \epsilon^{\pm} - h - \hbar \Sigma^{(\pm)} (\epsilon) \right] {\mathcal G}^{(\pm)} (\epsilon) = 1~,
\label{eq:RetAdvMag}
\end{equation}
where $\epsilon^\pm = \epsilon \pm i 0$.  The magnon self-energies have contributions from the leads, as well as a contribution from the bulk denoted by $\hbar \Sigma^{\rm FM}$:
\begin{equation}
\label{eq:sumretarded}
\hbar \Sigma (\epsilon) = \hbar  \Sigma^{{\rm FM}} (\epsilon) +\sum_{r \in \{L,R\}} \hbar \Sigma^{r} (\epsilon)~.
\end{equation}
From Eq.~(\ref{eq:keldyshsigmaleads}) we find that for the retarded and advanced component, the contribution due to the leads is given by
\begin{align}
\hbar \Sigma^{r, (\pm)}_{j,j'} (\epsilon) =  \sum_{k k' k'' k'''} J^r_{j;k k'} \left(J^r \right)^{*}_{j';k'' k'''} \int \frac{d\epsilon'}{(2\pi)}\int \frac{d\epsilon''}{(2\pi)}
\nonumber \\
 \times A_{k' k'';r} (\epsilon') A_{k''' k;r} (\epsilon'') \frac{N_F \left(\frac{\epsilon'-\mu_{r\uparrow}}{k_B T_r}\right)-N_F\left(\frac{\epsilon''-\mu_{r\downarrow}}{k_B T_r}\right)}{-\epsilon^{\pm}+\epsilon'-\epsilon'' }\nonumber \\~,
\end{align}
whereas the "lesser" self-energy can be shown to be of the form:
\begin{equation}\label{eq:lesserselfenergyleads}\hbar \Sigma^{r,<}_{j,j'} (\epsilon) = 2 i N_B \left(\frac{\epsilon-\Delta\mu_{r}}{k_B T_r}\right) {\rm Im}\left[ \hbar \Sigma^{r,(+)}_{j,j'} (\epsilon) \right]~,
\end{equation}
with $N_B(x) = [e^{x}-1]^{-1}$ the Bose-Einstein distribution function and $\Delta \mu_r = \mu_{r\uparrow}-\mu_{r\downarrow}$ the spin accumulation in lead $r$. 

Having established the contributions due to the leads, we consider the bulk self-energy $\hbar \Sigma^{\rm FM}$, which in principle could include various contributions, such as magnon conserving and nonconsering magnon-phonon interactions, or magnon-magnon interactions. 
Here, we consider magnon non-conserving magnon-phonon coupling as the source of the bulk self-energy and use the Gilbert damping phenomenology to parameterize it by the constant $\alpha$ which for the magnetic insulator YIG is of the order of $10^{-4}$. Gilbert damping corresponds to a decay of the magnons into phonons with a rate proportional to their energy. This thus leads to the contributions
\begin{eqnarray}
\label{eq:selfenergygilbert}
\hbar \Sigma^{{\rm FM},<}_{j,j'} (\epsilon)& = & 2  N_B \left(\frac{\epsilon}{k_B T_{\rm FM}}\right)  \hbar \Sigma^{{\rm FM},(+)}_{j,j'} (\epsilon) ~;~ \nonumber \\
 \hbar \Sigma^{{\rm FM},(+)}_{j,j'} (\epsilon) & = &  -i \alpha \, \epsilon \delta_{j,j'}~,
\end{eqnarray}
where $T_{\rm FM}$ is the temperature of phonon bath. We note that in principle the temperature could be taken position dependent to implement a temperature gradient, but we do not consider this situation here. 

With the results above, the density-matrix elements $\rho_{j,j'}$  can be explicitly computed from the magnon retarded and advanced Green's function and the ``lesser'' component of the total magnon self-energy using Eq.~(\ref{eq:dm_magnon}). The magnon self-energy is evaluated using the explicit expression for the retarded and advanced magnon self-energies due to leads and Gilbert damping $\hbar \Sigma^{{\rm FM}}$, see Eq.~(\ref{eq:selfenergygilbert}).

We are interested in the computation of the magnon spin current $\langle j_{m;j j'}\rangle$ in the bulk of the FM from site $j$ to site $j'$, which in terms of the magnon density matrix reads,  
\begin{equation}
\label{eq:spincurrentlocal}
\langle j_{m;j j'} \rangle= - i (h_{j,j'}\rho_{j',j} - {\rm c.c.})~,
\end{equation}
and follows from evaluating the change in time of the local spin density, $\hbar d\langle \hat b^\dagger_j \hat b_j \rangle/dt$, using the Heisenberg equations of motion. The magnon spin current in the bulk thus follows straightforwardly from the magnon density matrix. 

While the formalism presented so far provides a complete description of the magnon spin transport driven by metallic reservoirs, we discuss two simplifying developments below. First, we derive a Landauer-B\"utikker-like formula for the spin current from metallic reservoirs to the magnon system. Second, we discuss how to replace the matrix elements $J^r_{j;k,k'}$ by a single phenomenological parameter that characterizes the interface between metallic reservoirs and the magnetic insulator. 

\subsection{Landauer-B\"uttiker formula}
\label{landauer}
In this section we derive a Landauer-B\"uttiker formula for the magnon transport. Using the Heisenberg equations of motion for the local spin density, we find that the spin current from the left reservoir into the magnon system is given by
\begin{eqnarray}
j^L_s \equiv -\frac{\hbar}{2}\left\langle \frac{d}{dt}\sum_k\left(\hat \psi^\dagger_{k \uparrow L}\psi_{k \uparrow L}-\psi^\dagger_{k \downarrow L}\psi_{k \downarrow L}\right) \right\rangle   \nonumber \\=  -\frac{2}{\hbar} \sum_{j;k k^{\prime}}  {\rm Re}[\left(J^L\right)_{j;k k^{\prime}}^{*}
g^{<}_{j;k k^{\prime}}(t,t^{\prime})]~,
\end{eqnarray}
in terms of the Green's function
\begin{equation}
g^{<}_{j;k k^{\prime}}(t,t^{\prime})  \equiv i \langle \hat{\psi}^{\dagger}_{k^{\prime}  \uparrow L}(t')\hat{\psi}_{k \downarrow L}(t')\hat{b}_{j}(t)  \rangle  
\label{eq:interface_derivation}~.
\end{equation}
This ``lesser'' coupling Green's function $g^{<}_{j;k k^{\prime}}(t,t^{\prime})$ is calculated using Wick's theorem and standard Keldysh methods as described below.
  
We introduce the spin-flip operator for lead $r$
\begin{equation}
\hat{d}_{k k^{\prime};r}^{\dagger}(t) = \hat{\psi}^{\dagger}_{k^{\prime}  \uparrow r}(t)\hat{\psi}_{k \downarrow r}(t)~,
\end{equation}
so that the coupling Green's function becomes  
\begin{eqnarray}
g^{<}_{j;k k^{\prime}}(t,t^{\prime}) \equiv i \langle \hat{d}_{k k^{\prime};L}^{\dagger}(t^{\prime})\hat{b}_{j}(t)  \rangle~.
\label{eq:coupling}
\end{eqnarray}
The Keldysh Green's function for the spin-flip operator is given by
\begin{eqnarray}
\Pi^r_{k k^{\prime} k^{\prime \prime} k^{\prime \prime \prime}}(t,t^{\prime}) = -i \langle T_{\mathcal C^{\infty}} \hat{d}_{k k^{\prime};r}(t) \hat{d}_{k^{\prime \prime} k^{\prime \prime \prime};r}^{\dagger}(t^{\prime}) \rangle~ 
\end{eqnarray}
and using Wick's theorem we find that 
\begin{eqnarray}
 \Pi_{k k^{\prime} k^{\prime \prime} k^{\prime \prime \prime}}^{r,>}(t,t^{\prime}) 
 &= & -i\,G_{k k^{\prime\prime \prime};r\downarrow}^{>}(t,t^{\prime})G_{k^{\prime} k^{\prime \prime};r\uparrow}^{<}(t^{\prime},t)~; \nonumber   \\ 
\Pi_{k k^{\prime} k^{\prime \prime} k^{\prime \prime \prime}}^{r,<}(t,t^{\prime}) 
& =& -i \,G_{k^{\prime} k^{\prime \prime};r\uparrow}^{>}(t^{\prime},t)   G_{k k^{\prime\prime \prime};r\downarrow}^{<}(t,t^{\prime})~;\nonumber \\ 
 \Pi_{k k^{\prime} k^{\prime \prime} k^{\prime \prime \prime}}^{r,(+)}(t,t^{\prime}) \nonumber \\ &=& -i \theta(t-t^{\prime})\Big[ G_{k k^{\prime\prime \prime};r\downarrow}^{>}(t,t^{\prime})G_{k^{\prime} k^{\prime \prime};r\uparrow}^{<}(t^{\prime},t) \nonumber \\
&& - G_{k^{\prime} k^{\prime \prime};r\uparrow}^{>}(t^{\prime},t)   G_{k k^{\prime\prime \prime};r\downarrow}^{<}(t,t^{\prime})\Big]~, 
\end{eqnarray}
where we used the definition for the electron Green's function in Eq.~(\ref{eq:electronkeldyshgreensfct}).

Applying the Langreth theorem \cite{maciejko2007introduction} and Fourier transforming, we write down the lesser coupling Green's function in terms of the spin-flip Green's function  and magnon Green's function
\begin{eqnarray}
 g^{<}_{j;k k^{\prime}}(\epsilon)  &=& \sum_{j^{\prime};k^{\prime \prime} k^{\prime \prime \prime}} J^L_{j';k'' k'''}  \Big( \mathcal{G}_{j,j^{\prime}}^{(+)}(\epsilon)\Pi_{k k^{\prime} k^{\prime \prime} k^{\prime \prime \prime}}^{L,<}(\epsilon) \nonumber \\ &&+\mathcal{G}_{j^{\prime},j}^{<}(\epsilon)\Pi_{k k^{\prime} k^{\prime \prime} k^{\prime \prime \prime}}^{L,(-)}(\epsilon) \Big)~,\label{eq:coupleGF}
\end{eqnarray}
where the retarded and ``lesser" magnon Green's function are given by Eq.~(\ref{eq:dm_magnon}) and Eq.~(\ref{eq:RetAdvMag}). Using these results, we ultimately find that 
\begin{eqnarray}
j_s^L &=& \int \frac{d \epsilon  }{2 \pi }   \, \left[N_{B}\left(\frac{\epsilon-\Delta\mu_{L}}{k_{B}T_{L}}\right)-N_{B}\left(\frac{\epsilon-\Delta\mu_{R}}{k_{B}T_{R}}\right)\right]{\rm T}(\epsilon) \nonumber \\
&+&\int \frac{d \epsilon  }{2 \pi }   \, \left[N_{B}\left(\frac{\epsilon-\Delta\mu_{L}}{k_{B}T_{L}}\right)-N_{B}\left(\frac{\epsilon}{k_{B}T_{\rm FM}}\right)\right] \nonumber \\
&\times& {\rm Tr} \left[
\hbar \Gamma^{L}(\epsilon)   {\mathcal G}^{(+)} (\epsilon)
\hbar \Gamma^{{\rm FM}}(\epsilon){\mathcal G}^{(-)} (\epsilon) \right]~,
\label{eq:lb_current}
\end{eqnarray}
with the transmission function
\begin{equation} 
\label{eq:transmission}
 {\rm T}(\epsilon) \equiv {\rm Tr} \left[
\hbar \Gamma^{L}(\epsilon)   {\mathcal G}^{(+)} (\epsilon)
\hbar \Gamma^{R}(\epsilon){\mathcal G}^{(-)} (\epsilon) \right]~.
\end{equation} 
In the above, the rates $ \hbar \Gamma^{L/R}(\epsilon)$ are defined by
\begin{equation}
\label{eq:ratesinterface}
 \hbar \Gamma^{r}(\epsilon) \equiv -2 {\rm Im}\left[\hbar \Sigma^{r,(+)}(\epsilon)\right]~,
\end{equation}
and 
\begin{equation}
\hbar \Gamma^{\rm FM}(\epsilon) \equiv -2 {\rm Im}\left[\hbar \Sigma^{\rm FM,(+)}(\epsilon)\right]~,
\end{equation}
and correspond to the decay rates of magnons with energy $\epsilon$ due to interactions with electrons in the normal metal at the interfaces, and phonons in the bulk, respectively. This result is similar to the Laudauer-B\"uttiker formalism \cite{di2008electrical} for electronic transport using single-particle scattering theory. In the present context, a Landauer-B\"uttiker-like for spin transport was first derived by Bender {\it et al.} [\onlinecite{bender2012electronic}] for a single NM-FM interface.  
In the absence of Gilbert damping, the spin current would correspond to the expected result from Landauer-B\"utikker theory, i.e., the spin current from left to the right lead is then given by the first line of Eq.~(\ref{eq:lb_current}). The presence of damping gives leakage of spin current due to the coupling with the phononic reservoir, as the second term shows.
Finally, we note that the spin current from the right reservoir into the system
is obtained by interchanging labels L and R in the first term, and the label L replaced by R in the second one. 
Due to the presence of Gilbert damping, however, we have in general that $j_s^L \neq -j_s^R$.

\subsection{Determining the interface coupling}
We now proceed to express the magnon spin current (Eq.~(\ref{eq:lb_current})) in terms of a macroscopic, measurable quantity rather than the interfacial exchange constants $J^r_{j;k,k'}$. For $\Delta\mu_{r} \ll \epsilon_F$ (with $\epsilon_F$ the Fermi energy of the metallic leads), which is in practice always obeyed, we have for low energies and temperatures that
\begin{eqnarray}
\label{eq:selfenergyohmic}
& \hbar \Sigma^{r,(\pm)}_{j,j'} (\epsilon) &
\simeq  \mp i \frac{1}{4\pi} \sum_{k k' k'' k'''} J^r_{j;k k'} \left(J^r\right)^{*}_{j';k'' k'''}
\nonumber \\
&& A_{k',k'';r} (\epsilon_F) A_{k''',k;r} (\epsilon_F) (\epsilon-\Delta \mu_r)~.
\end{eqnarray}
Here, we also neglected the real part of this self-energy which provides a small renormalization of the magnon energies but is otherwise unimportant. The expansion for small energies in Eq.~(\ref{eq:selfenergyohmic}) is valid as long as $\epsilon \ll \epsilon_F$, which applies since $\epsilon$ is a magnon energy, and therefore at most on the order of the thermal energy.
Typically, the above self-energy is strongly peaked for $j,j'$ at the interface because the magnon-electron interactions occur at the interface. For $j,j'$ at the interface we have that the self-energy depends weakly on varying $j,j'$ along the interface provided that the properties of the interface do not vary substantially from position to position. We can thus make the identification:
\begin{eqnarray}
&&\hbar \Sigma^{r, (\pm)}_{j,j'} (\epsilon) 
\simeq  \mp i \eta^r (\epsilon-\Delta \mu_r)\delta_{j,j'}\delta_{j,j_r}~,
\label{eq:self_energy}
\end{eqnarray}
with $j_r$ the positions of the sites at the $r$-th interface, and $\eta^r$ parametrizing the coupling between electrons and magnons at the interface. Note that $\eta^r$ can be read off from Eq.~(\ref{eq:selfenergyohmic}). Rather than evaluating this parameter in terms of the matrix elements $J^r_{j;k k'}$ and the electronic spectral functions of the leads $A_{k,k';r}(\epsilon)$, we determine it in terms of the real part of the spin-mixing conductance $g^{\uparrow \downarrow;r}$, a phenomenological parameter that characterizes the spin-transfer efficiency at the interface \cite{brataas2000finite}. This can be done by noting that in the classical limit the self-energy in Eq.~(\ref{eq:self_energy}) leads to an interfacial contribution, determined by the damping constant $\eta^r/N$, to the Gilbert damping of the homogeneous mode, 
 where $N$ is the number of sites of the system perpendicular to the leads, as indicated in Fig.~\ref{fig:schematic}. 
In terms of the spin-mixing conductance, we have that this contribution is given by \cite{tserkovnyak2002enhanced} $g^{\uparrow \downarrow;r}/4 \pi s_r  N$, with $s_r$ the saturation spin density per area of the ferromagnet at the interface with the $r$-th lead. Hence, we find that  
\begin{equation}
\label{eq:eta}
\eta^r  =\frac{g^{\uparrow \downarrow;r}}{4\pi s_r}~,
\end{equation}
which is used to express the reservoir contributions to the magnon self-energies in terms of measurable quantities. The spin-mixing conductance can be up to $5 \hbar$ nm$^{-2}$ for YIG-Pt interfaces \cite{jia2011}, leading to the conclusion that $\eta$ can be of 
the order $1-10$ for that case.  

\subsection{Summary on implementation} \label{sec:summary}
We end this section with some summarizing remarks on implementation that may facilitate the reader who is interested in applying the formalism presented here.

First, one determines the retarded and advanced magnon Green's functions. This can be done given a magnon hamiltonian characterized by matrix elements $h_{j,j'}$ in Eq.~(\ref{eq:magnonham}), mixing conductances for the metal-insulator interfaces $g^{\uparrow\downarrow; r}$, and a value for the Gilbert damping constant $\alpha$, from which one computes the retarded self-energies at the interfaces in Eq.~(\ref{eq:self_energy}) with Eq.~(\ref{eq:eta}), and Eq.~(\ref{eq:selfenergygilbert}). The retarded and advanced magnon Green's functions are then computed via Eq.~(\ref{eq:RetAdvMag}), which amounts to a matrix inversion. The next step is to calculate the density matrix for the magnons using Eq.~(\ref{eq:dm_magnon}), with as input the expressions for the ``lesser'' self-energies in Eqs.~(\ref{eq:lesserselfenergyleads})~and~(\ref{eq:selfenergygilbert}). Finally, the spin current is evaluated using Eq.~(\ref{eq:spincurrentlocal}) in the bulk of the FM or Eq.~(\ref{eq:lb_current}) at the NM-FM interface. In the next sections, we discuss some applications of our formalism.

\section{Numerical results}
\label{numerical}

In this section, we present results of numerical calculations using the formalism presented in the previous section. 
\begin{table}[]
	\centering
	\caption{Parameters chosen for numerical calculations based on the NEGF formalism (unless otherwise noted).\\
	}
	\label{parameters}
	\begin{tabular}{|l|l|l|l|l|} \hline
		& Quantity   & Value \\ \hline
		& $J$ &  $0.05 \,eV$  \\
		& $\Delta\mu_{L}/J$ & $ 2.0 \times 10^{-5} $    \\
		& $\Delta\mu_{R}/J $ & $ 0.0 $ \\
		&  $\eta$ & $8$    \\
		&  $ \Delta/J$ & $2.0 \times 10^{-3}$  \\
		&  $k_BT_{\rm FM}/J$  & $ 0.60  $ \\ 
		\hline
	\end{tabular}
\end{table}

\subsection{Clean system}
For simplicity, we consider now the situation where the leads and magnetic insulators are one dimensional. The values of various parameters are displayed in Table \ref{parameters}, where we take the hopping amplitudes $J_{j,j'}=J(\delta_{j,j'+1}+\delta_{j,j'-1})$, i.e., $J_{j,j'}$ is equal to $J$ between nearest neighbours, and zero otherwise. 
 We focus on transport driven by spin accumulation in the leads and set all temperatures equal, i.e., $ T_{L}=T_{R}=T_{\rm FM} \equiv T$. We also assume both interfaces to have equal properties, i.e., for the magnon-electron coupling parameters to obey $\eta^L=\eta^R\equiv \eta$. First we consider the case without disorder and take $\Delta_j=\Delta$.
 
\begin{figure}[!htb]
	\includegraphics[scale=0.42]{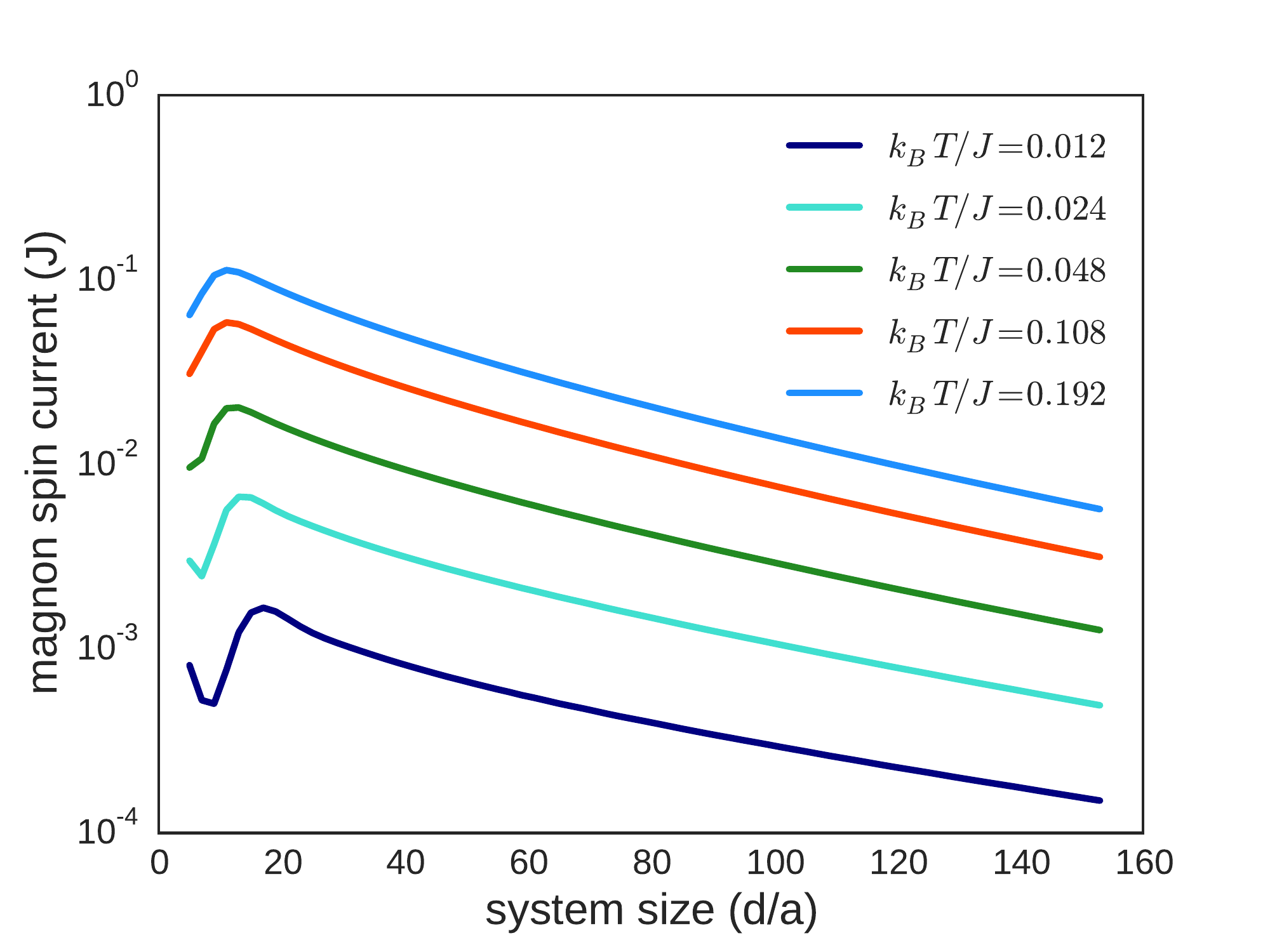}\caption{\small System-size dependence of spin current ejected in the right reservoir for 
		$\alpha=6.9\times 10^{-2},\eta = 8.0$ and various temperatures. }
	\label{fig:magnonEjectedLog}
\end{figure}

\begin{figure}[!htb]
	\includegraphics[scale=0.35]{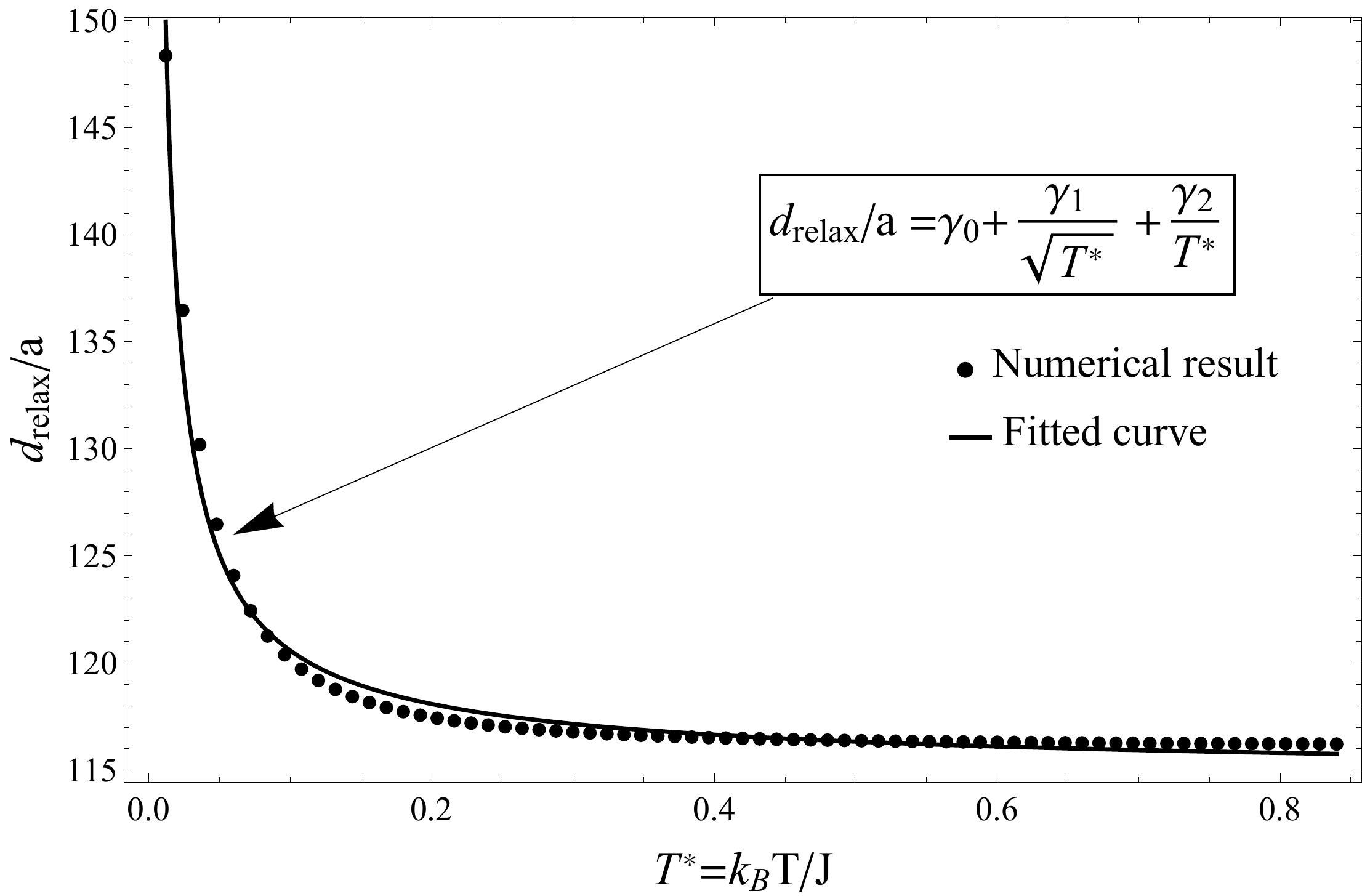}\caption{\small Magnon relaxation length as a function of dimensionless temperature $T^{*}$ for $\alpha=6.9\times 10^{-2}, \eta=8.0$. The fitted parameters are obtained as $\gamma_{0}=114.33,\gamma_{1}= 0.96,\gamma_{2} = 0.32$.}
	\label{fig:fit_relaxation}
\end{figure}

  We are interested in how the Gilbert damping affects the magnon spin current. In particular, we calculate the spin current injected in the right reservoir as a function of system size. The results of this calculation are shown in Fig.~\ref{fig:magnonEjectedLog} for various temperatures, which indicates that for a certain fixed spin accumulation, the injected spin current decays with 
 the thickness of the system for $N>25$, for the parameters we have chosen. We come back to the various regimes of thickness dependence when we present analytical results for clean systems in the continuum limit in Sec.~\ref{sec:analytical}. 
From these results we define a magnon relaxation length $d_{\rm relax}$ using the definition 
\begin{equation}
j_{m}(d) \propto \exp(-d/d_{\rm relax})~,
\end{equation}
applied to the region $N>25$ and where $d={\rm N}a$ with $a$ the lattice constant. The magnon relaxation length depends on system temperature and is shown in Fig.~\ref{fig:fit_relaxation}. We attempt to fit the temperature dependence with 
\begin{equation}
d_{\rm relax} (T^{*})=a ( \gamma_{0}+\frac{\gamma_{1}}{\sqrt{T^{*}}}+\frac{\gamma_{2}}{T^{*}})~,
\end{equation}
with $\gamma_0, \gamma_1, \gamma_2$ constants and $T^{*}$ defined as the dimensionless temperature $T^{*}\equiv k_{B}T/J$. The term proportional to $\gamma_1$ is expected for quadratically dispersing magnons with Gilbert damping as the only relaxation mechanism \cite{hoffman2013landau,cornelissen2015long}. The terms proportional to $\gamma_0$ and $\gamma_2$ are added to characterize the deviation from this expected form. Our results show that the relaxation length has not only $\sim 1/\sqrt{T}$ behaviour. This is due to the finite system size,  the contact resistance that the spin current experiences at the interface between metal and magnetic insulator, and the deviation of the magnon dispersion from a quadratic one due to the presence of the lattice.

\subsection{Disordered system}
 We now consider the effects of disorder on the spin current as a function of the thickness of the FM. We consider a one-dimensional system with a disorder potential implemented by taking $\Delta_j=\Delta(1+\delta_j)$, 
 where $\delta_j$ is a random number evenly distributed between $-\delta$ and $\delta$ (with $\delta \ll 1$ and positive) that is uncorrelated between different sites. In one dimension, all magnon states are Anderson localized \cite{PhysRev.109.1492}. Since this is an interference phenomenon, it is expected that Gilbert damping diminishes such localization effects. The effect of disorder on spin waves was investigated using a classical model in Ref.~[\onlinecite{PhysRevB.92.014411}], whereas Ref.~[\onlinecite{PhysRevLett.118.070402}] presents a general discussion of the effect of dissipation on Anderson localization. Very recently, the effect of Dzyaloshinskii-Moriya interactions on magnon localization was studied \cite{evers2017}. Here we consider how the interplay between Gilbert damping and the disorder affects the magnon transport.
 
\begin{figure}[!htb]
 \hspace*{-1.98em}	\includegraphics[scale=0.33]{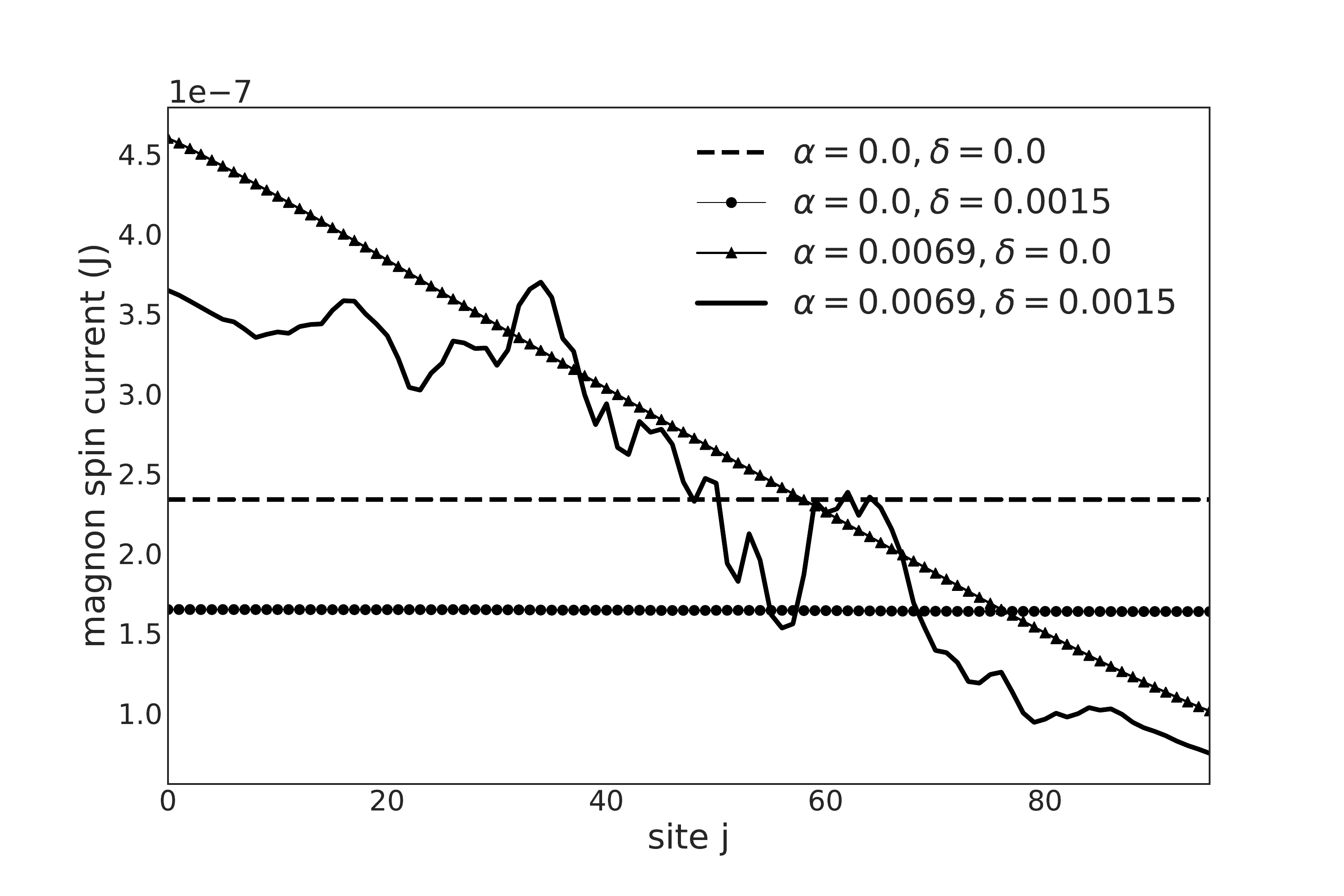}\caption{\small Spatial dependence of local magnon current for the case without Gilbert damping and disorder ($\alpha=0, \delta=0$), without disorder ($\alpha=6.9\times 10^{-3}, \delta=0$), without Gilbert damping ($\alpha=0,\delta=1.5\times 10^{-3}$), and both disorder and Gilbert damping ($\alpha=6.9\times 10^{-3},\delta=1.5\times 10^{-3}$). The interface coupling parameter is taken equal to $\eta=0.8$.} 
\label{fig:localmagnoncurrentdampingdisorder}
\end{figure}

For a system without Gilbert damping the spin current carried by magnons is conserved and therefore independent of position regardless of the presence or absence of disorder. Due to the presence of Gilbert damping the spin current decays as a function of position. Adding disorder on top of the dissipation due to Gilbert damping causes the spin current to fluctuate from position to position. For large Gilbert damping, however, the effects of disorder are suppressed as the Gilbert damping suppresses the localization of magnon states. In Fig.~\ref{fig:localmagnoncurrentdampingdisorder} we show numerical results of the position dependence of the magnon current 
for different combinations of disorder  and Gilbert damping constants.
 The plots clearly show that the spin current fluctuates in position due to the combined effect of disorder and Gilbert damping, whereas it is constant without Gilbert damping, and decays in the case with damping but without disorder. Note that for the two cases without Gilbert damping the magnitude of the spin current is different because the disorder alters the conductance of the system and each curve in Fig.~\ref{fig:localmagnoncurrentdampingdisorder} corresponds to a different realization of disorder. 

\begin{figure}[!htb]
	\includegraphics[scale=0.30]{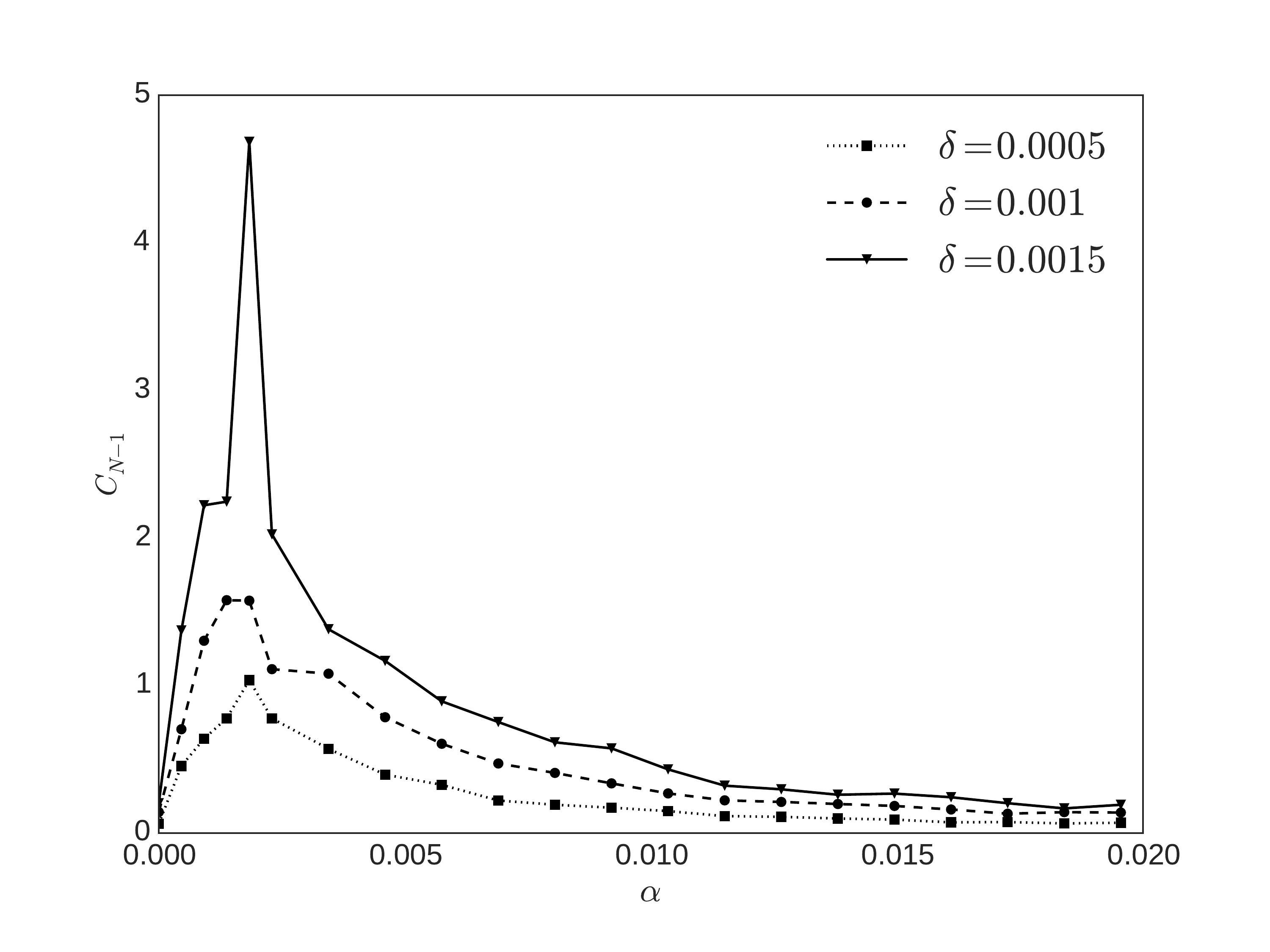}\caption{\small Correlation function $C_j$  that characterizes the fluctuations in the spin curent for $j=N-1$ as a function of the Gilbert damping constant, for three strengths of the disorder potential. The curves are obtained by performing averaging over 100 realizations of the disorder. The interface coupling parameter is taken equal to $\eta=0.8$.}
	\label{fig:flucsspincurrent}
\end{figure}

To characterize the fluctuations in the spin current, we define the correlation function
\begin{equation}
  C_{j} = \sqrt{ \frac{\overline{\left(j_{m;j,j+1}-\overline{j_{m;j,j+1}}\right)^2}}{\left(\overline{j_{m;j,j+1}}\right)^2}}~,
\end{equation}
where the bar stands for performing averaging over the realizations of disorder. 
Fig.~(\ref{fig:flucsspincurrent}) shows this correlation function for $j=N-1$ as a function of Gilbert damping for various strengths of the disorder. As we expect, based on the previous discussion, the fluctuations become small as the Gilbert damping becomes very large or zero, leaving an intermediate range where there are sizeable fluctuations in the spin current.  

\section{Analytical results}
\label{sec:analytical}
 In this section we analytically compute the magnon transmission function in the continuum limit $a\rightarrow 0$ for a clean system. We consider again the situation of a magnon hopping amplitude $J_{j,j'}$ that is equal to $J$ and nonzero only for nearest neighbors, and a constant magnon gap $\Delta_j=\Delta$.
We compute the magnon density matrix, denoted by $\rho (\bx,\bx')$, and retarded and advanced Green's functions, denoted by $\mathcal{G}^{(\pm)}(\bx, \bx^{\prime \prime};\epsilon)$. 
 Here, the spatial coordinates in the continuum are denoted by $\bx,\bx',\bx'', \cdots$. 
 We take the system to be translationally invariant in the $y-z$-plane and the current direction as shown in Fig.~\ref{fig:device} to be $x$. 

In the continuum limit, the imaginary part of the various self-energies acquired by the magnons have the form:
\begin{eqnarray}
	\label{eq:selfenergiescontinuum}
	&&\mathrm{Im}\left[\hbar \Sigma^{r,(+)}(\bx^,\bx^{\prime};\epsilon)\right] = \nonumber \\
	&& - \tilde \eta^r(\epsilon-\mu_{r})\delta(x-x_{r})\delta(\bx-\bx')~;\nonumber \\
	&& \mathrm{Im}\left[\hbar \Sigma^{{\rm FM},(+)}(\bx,\bx^{\prime };\epsilon)\right] = - \alpha \,\epsilon\, \delta(\bx-\bx^{\prime})\, ,
\end{eqnarray}
where $x_r$ is the position of the $r$-th lead, and where $\tilde \eta^r$ is the parameter that characterizes the interfacial coupling between magnons and electrons. We use a different notation for this parameter as in the continuum situation its dimension is different with respect to the discrete case. To express $\tilde \eta^r$ in terms of the spin-mixing conductance we have that $\tilde \eta^r=g^{\uparrow\downarrow}/4\pi \tilde s^r$ where $\tilde s^r$ is now the three-dimensional saturated spin density of the ferromagnet. 

We proceed by evaluating the magnon transmission function from Eq.~(\ref{eq:transmission}). We compute the rates in Eq.~(\ref{eq:ratesinterface}) from the self-energies Eqs.~(\ref{eq:selfenergiescontinuum}) and find for the transmission function in the first instance that
\begin{eqnarray}
\label{eq:trans1st}
	\label{eq:transmissionfunction} 
	&& {\rm T}(\epsilon) = 4 \tilde \eta^L \tilde \eta^R (\epsilon-\Delta\mu_L) (\epsilon-\Delta\mu_R) \nonumber \\ && \times \int \frac{d\bq}{(2\pi)^2} g^{(+)} (x_L,x_R,\bq;\epsilon) g^{(-)} (x_R,x_L,\bq;\epsilon)~,
\end{eqnarray}
where $\bq$ is the two-dimensional momentum that results from Fourier transforming in the $y-z$-plane. The Green's functions $g^{(\pm)} (x,x',\bq;\epsilon)$ obey [compare Eq.~(\ref{eq:RetAdvMag})]
\begin{eqnarray}
	\label{eq:gfeqcontinuum} 
	&& \left[ (1\pm i \alpha)\epsilon +A\frac{d^2}{dx^2}-A\bq^2 -\Delta \right. \nonumber \\ 
&&	\left. \pm i \sum_{r\in\{L,R\}} \tilde \eta^{r} (\epsilon-\Delta \mu_r) \delta (x-x_r)\right] g^{(\pm)} (x,x',\bq;\epsilon)
	\nonumber \\
	&&=\delta (x-x')~,
\end{eqnarray}
where $A=Ja^2$.  
This Green's function is evaluated using standard techniques for inhomogeneous boundary value problems (see Appendix \ref{app:greensfct}) to ultimately yield
\begin{equation}
	\label{eq:transNEGF}
	{\rm T}(\epsilon) = 4 \tilde \eta^2  (\epsilon-\Delta\mu_L) (\epsilon-\Delta\mu_R)  \int \frac{d\bq}{(2\pi)^2} |t(\bq,\epsilon)|^2~,
\end{equation}
with 
\begin{eqnarray}
\label{eq:tqe}
	&& t (\bq, \epsilon) = A  \kappa  \left[\left(A^2  \kappa^2 - \tilde \eta^2 (\epsilon- \Delta\mu_L)(\epsilon-\Delta\mu_R) \right)\sinh ( \kappa d)\right. \nonumber \\
	&& \left.-i A \tilde \eta  \kappa (2\epsilon-\Delta\mu_L-\Delta\mu_R) \cosh ( \kappa d)\right]^{-1},
\end{eqnarray}
with $ \kappa = \sqrt{(A\bq^2 + \Delta-\epsilon-i\alpha\epsilon)/A}$ and where $d=x_R-x_L$. Note that we have at this point taken both interfaces equal for simplicity, so that $\tilde \eta^L=\tilde \eta^R \equiv \tilde \eta$. In terms of an interfacial Gilbert damping parameter $\alpha'$ we have that $\tilde \eta = d \alpha'$.  

Let us identify the magnon decay length 
\[
l\equiv\frac{\lambda}{\alpha},
\]
where $\lambda=\sqrt{A/k_B T}$  is proportional to the thermal de Broglie wavelength.
Equipped with a closed, analytic expression, we may now, in an analogous way as Hoffman {\it et al.}~[\onlinecite{hoffman2013landau}], investigate the behavior of Eq.~(\ref{eq:transNEGF}) in the thin FM ($d\ll l$) and thick FM ($d\gg l$) regimes. In order to do so, we take $\mu_L=0$ so that the second term in Eq.~(\ref{eq:lb_current}) vanishes and the spin current is fully determined by the transmission coefficient ${\rm T} (\epsilon)$. Before analyzing the result for the spin current more closely, we remark that the result for the transmission function may also be obtained from the linearized stochastic Landau-Lifshitz-Gilbert equation, as shown in Appendix~\ref{app:stochastic}.

\subsection{Thin film regime ($d \ll l$)}
\label{thin}

In the thin film regime, the transmission coefficient ${\rm T} (\epsilon)$ exhibits scattering resonances near $\epsilon=\epsilon_{n\mathbf{q}}$
for given $\mathbf{q}$, where
\[
\frac{\epsilon_{n\mathbf{q}}}{A}=\mathbf{q}^{2}+\frac{1}{\xi^{2}}+\frac{n^{2}\pi^{2}}{d^{2}}
\]
and $n$ is an integer and where $\xi=\sqrt{A/\Delta}$ is the coherence length of the ferromagnet. When the ferromagnet is sufficiently thin ($d\ll\lambda/\alpha^{1/2}=\sqrt{\alpha}l$),
one finds that these peaks are well separated, and the transmission coefficient is approximated as a sum of Lorentzians: ${\rm T} =\sum_{n=0}^\infty {\rm T}_n$, where:
\begin{equation}
{\rm T}_n(\epsilon)\approx A_{n\mathbf{q}} \frac{\Gamma^L_n\Gamma^R_n}{\Gamma^L_n+\Gamma^R_n+\Gamma^{\rm{FM}}_n}
\end{equation}
with
\begin{equation}
A_{n\mathbf{q}} (\epsilon )=\frac{\Gamma_n}{\left(\epsilon-\epsilon_{n\mathbf{q}} \right)^2+\left(\Gamma_n/2 \right)^2}\, 
\end{equation}
as the spin wave spectral density. The broadening rates are given by $\Gamma_n^{\rm{FM}}=2\alpha \epsilon$, $\Gamma_0^L=2\alpha '\epsilon$, $\Gamma_0^R=2\alpha '(\epsilon-\mu_R)$, $\Gamma_{n \neq 0}^L=4\alpha '\epsilon$, $\Gamma_{n \neq0}^R=4\alpha '(\epsilon-\mu_R)$ and $\Gamma_n=\Gamma_n^{\rm{FM}}+\Gamma_n^L+\Gamma_n^R$.  In the extreme small dissipation limit (i.e.
neglecting spectral broadening by the Gilbert damping), one has: 
\begin{equation}
A_{n\mathbf{q}}\left(\epsilon\right)\rightarrow2\pi\delta\left(\epsilon-\epsilon_{n{\bf q}}\right)\, ,
\end{equation}
and the current has the simple form, $j_{s}^L=\sum_{n=0}^{\infty}j_n$, where
\begin{equation}
j_n=a^2 \int \frac{d^2 q}{(2\pi)^2}  \frac{\Gamma^L_n\Gamma^R_n}{\Gamma_n}\left[N_{B}\left(\frac{\epsilon_{n\mathbf{q}}}{k_{B}T}\right)-N_{B}\left(\frac{\epsilon_{n\mathbf{q}}-\Delta\mu_{R}}{k_{B}T}\right)\right]
\label{jclsmall}
\end{equation}
where $\Gamma^L_n$, $\Gamma^R_n$ and $\Gamma^{\rm{FM}}_n$ are all evaluated at $\epsilon=\epsilon_{n\mathbf{q}}$. Eq.~(\ref{jclsmall}) allows one to estimate the thickness dependence of the signal. Supposing $\mu_R \lesssim \epsilon_{n\mathbf{q}}$, when $d \ll g^{\uparrow \downarrow}/s\alpha $, then $\alpha' \gg \alpha$, and  $ \Gamma^L_n\Gamma^R_n/\Gamma^{\rm{FM}}_n\sim j_{s,\rm{cl}}^L \sim 1/d$; when $d \gg g^{\uparrow \downarrow}/s\alpha $, then $\alpha' \ll \alpha$, and  $j_{s,\rm{cl}}^L \sim 1/d^2$. The enhancement of the spin current for small $d$ is in rough agreement with our numerical results in the previous section as shown in Fig.~\ref{fig:magnonEjectedLog}.

\subsection{Thick film regime ($d \gg l$) }
\label{thick}

In the thick film regime, the transmission function becomes
\begin{equation}
{\rm T}(\epsilon)\approx \frac{(4 A d)^2 \Gamma^L_x \Gamma^R_x\sqrt{(\epsilon-\epsilon_{0\mathbf{q}})^2+(\Gamma^{\rm{FM}}_x/2)^2}e^{-2\kappa_r d}}{\left| (4A\kappa)^2-(d)^2 \Gamma^L_x \Gamma^R_x -i4dA \Gamma^R_x \kappa S(\kappa_r)\right|^2}
\nonumber
\end{equation}
where $\Gamma^{L/R/\rm{FM}}_x=\Gamma^{L/R/\rm{FM}}_{n\neq0}$,
$\kappa_{r} = {\rm Re}[\kappa]$, 
and $S\left(\kappa_{r}\right)$
is the sign of $\kappa_{r}$. For $\alpha\ll1$, we have $\kappa=ik_{x}\left(1+i\alpha\epsilon/2A k_{x}^{2}\right)$,
where $k_{x}=\sqrt{\mathbf{q}^{2}+\xi^{-2}-\epsilon/A}$. For energies $\epsilon>A(\mathbf{q}^2+\xi^{-2})$, $k_{x}$ is imaginary, and the contribution
to the spin current decays rapidly with $d$. When, however,  $\epsilon<A(\mathbf{q}^2+\xi^{-2})$,
$k_{x}$ is real, and $\kappa_{r}=-\alpha\left(\mathbf{q}^{2}+\xi^{-2}\right)/2k_{x}\sim\alpha/\lambda$
(for thermal magnons), so that the signal decays over a length scale
$l\propto 1/\sqrt{T}$, in agreement with our numerical results as shown in Fig.~\ref{fig:fit_relaxation}. 

\subsection{Comparison with numerical results}
\begin{figure}[h!]
	\begin{center}
		\includegraphics[scale=0.45]{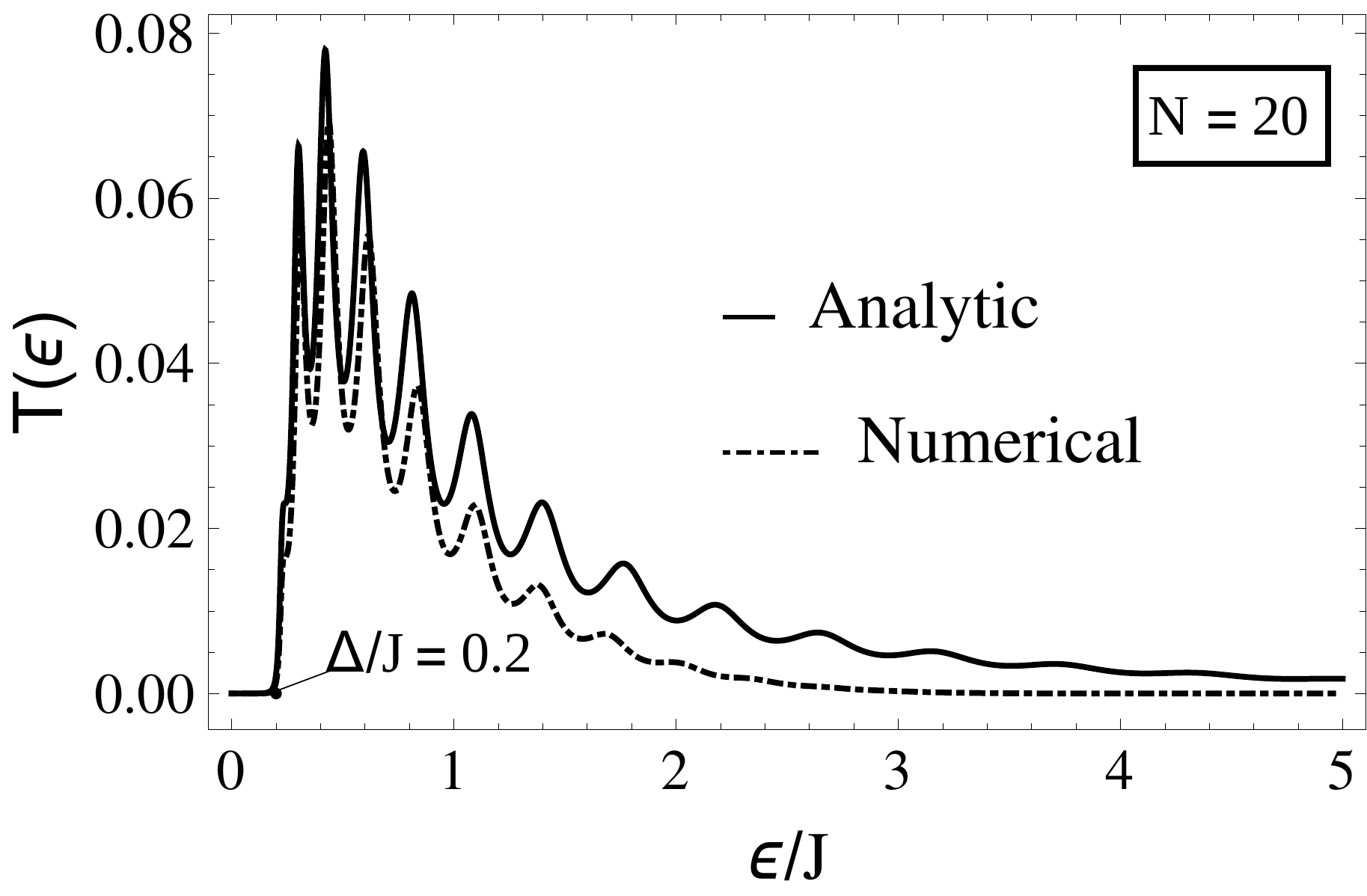}
		\caption{Magnon transmission function as a function of energy. The parameters are chosen to be $\Delta/J=0.2, \alpha=0.069, \eta=8.0$. }
		\label{fig:CompareTransmission}
	\end{center}
\end{figure}

In order to compare the numerical with the analytical results we plot in Fig.~\ref{fig:CompareTransmission} the transmission function as a function of energy. Here, the numerical result is evaluated for a clean system using Eq.~(\ref{eq:transmission}) while the analytical result is that of Eq.~(\ref{eq:transNEGF}). While they agree in the appropriate limit ($N \to \infty, a \to 0$), for finite $N$ there are substantial deviations that are due to the increased importance of interfacing coupling relative to the Gilbert damping for small systems and the deviations of the dispersion from a quadratic one. 

\section{Discussion and Outlook}
\label{discussion}
We have developed a NEGF formalism for exchange magnon transport in a NM-FM-NM heterostructure. We have illustrated the formalism with numerical and analytical calculations and determined the thickness dependence of the magnon spin current. We  have also considered magnon disorder scattering and shown that the interplay between disorder and Gilbert damping leads to spin-current fluctuations. 

We have also demonstrated that for a clean system, i.e., without disorder, in the continuum limit the results obtained from the NEGF formalism agree with those from the stochastic LLG formalism. The latter is suitable for a clean system in the continuum limit where the various boundary conditions on the solutions of the stochastic equations are easily imposed. The NEGF formalism is geared towards real-space implementation, such that, e.g., disorder scattering due to impurities are more straightforwardly included as illustrated by our example application. The NEGF formalism is also more flexible for systematically including self-energies due to additional physical processes, such as magnon-conserving magnon-phonon scattering and magnon-magnon scattering, or, for example, for treating strong-coupling regimes into which the stochastic Landau-Lifshitz-Gilbert formalism has no natural extension.

Using our formalism, a variety of mesoscopic transport features of magnon transport can be investigated including, e.g., magnon shot noise
  \cite{kamra2016magnon}. The generalization of our formalism to elliptical magnons and magnons in antiferromagnets is an attractive direction for future research.

\acknowledgements This work was supported by the Stichting voor Fundamenteel Onderzoek der Materie (FOM), the Netherlands
Organization for Scientific Research (NWO), and by the
European Research Council (ERC) under the Seventh
Framework Program (FP7). J.~Z. would like to thank the China Scholarship Council. J.~A. has received funding from the European Union's Horizon 2020 research and innovation programme under 
the Marie Sk\l odowska-Curie grant agreement No 706839 (SPINSOCS).

\begin{widetext}
\appendix
\section{Evaluation of magnon Green's function in the continuum limit} \label{app:greensfct}

In this appendix we evaluate the magnon Green's function in the continuum limit that is determined by Eq.~(\ref{eq:gfeqcontinuum}). For simplicity we take the momentum $\bq$ equal to zero and suppress it in the notation, as it can be trivially restored afterwards. The Green's function is then determined by 
\begin{equation}
\left[ \epsilon \pm i \alpha \epsilon \pm i \sum_{r \in \{L,R\}} \tilde \eta^{r}
(\epsilon-\Delta\mu_{r})\delta(x-x_{r}) + A \frac{d^2}{d x^2} - H \right] g^{(\pm)}(x,x';\epsilon)
= \delta(x-x')~.
\end{equation}
To determine this Green's function we first solve for the states $\chi^{\pm}(x)$ that obey:
\begin{equation}
\left[ \epsilon \pm i \alpha \epsilon \pm i \sum_{r \in \{L,R\}} \tilde \eta^{r}
(\epsilon-\Delta\mu_{r})\delta(x-x_{r}) + A \frac{d^2}{d x^2} - H \right] \chi^{\pm}(x)
= 0~.
\end{equation}
Integrating this equation across $x=x_{L}$ and  $x=x_{R}$ leads to the boundary conditions:
\begin{eqnarray}
x=x_{L} &:&\quad \pm i  \tilde \eta^{L} (\epsilon-\Delta\mu_{L})\chi^{\pm}(x_{L}) + A \frac{d \chi^{\pm}(x) }{d x}|_{x=x_{L}}  =0; \\
\label{eq:bound1}
x=x_{R} &:&\quad \pm i  \tilde \eta^{R} (\epsilon-\Delta\mu_{R})\chi^{\pm}(x_{R}) - A \frac{d \chi^{\pm}(x) }{d x}|_{x=x_{R}}   =0~.
\label{eq:bound2}
\end{eqnarray}
For $x_{L}<x<x_{R}$, the general solution is:
\begin{equation}
\chi^{\pm}(x) = B e^{i k_{\pm} x} + C  e^{-i k_{\pm} x}~,
\end{equation}
with $k_{\pm}=\sqrt{(\epsilon \pm i \alpha \epsilon-H)/A}$.
We write the solution obeying the boundary condition at $x=x_{L}$ as 
\begin{equation}
\chi_{L}^{\pm}(x) = e^{i k_{\pm} x} + C_{\pm}  e^{-i k_{\pm} x}~,
\end{equation}
With the boundary condition at $x=x_{L}$ (~Eq.\ref{eq:bound1}), we find that 
\[ 
C_{\pm} = \left[ \frac{A k_{\pm} \pm \tilde \eta^{L}(\epsilon-\Delta\mu_{L})}{A k_{\pm} \mp \tilde \eta^{L}(\epsilon-\Delta\mu_{L})}  \right]e^{2 i k_{\pm}x_{L}}~.
\]
For the solution obeying the boundary condition at $x=x_{R}$, we write 
\begin{equation}
\chi_{R}^{\pm}(x) = B_{\pm} e^{i k_{\pm} x} +  e^{-i k_{\pm} x}~.
\end{equation}
With the boundary condition at $x=x_{R}$ (~Eq.\ref{eq:bound2}), we
have:
	\[ 
	A(i B_{\pm}k_{\pm}e^{i k_{\pm} x_{R}} - i k_{\pm} e^{-i k_{\pm} x_{R}} ) = \pm i  \tilde \eta^{R}(\epsilon-\mu_{R})(B_{\pm} e^{i k_{\pm} x_{R}} +  e^{-i k_{\pm} x_{R}})~,
	\]
so that
\[ 
B_{\pm} = \left[ \frac{A k_{\pm} {\pm} \tilde \eta^{R}(\epsilon-\Delta\mu_{R})}{A k_{\pm} { \mp} \tilde \eta^{R}(\epsilon-\Delta\mu_{R})}  \right]e^{-2 i k_{\pm}x_{R}}~.
\]
The Green's function is now given by \cite{hermanbook}
\begin{equation}
g^{(\pm)}(x,x';\epsilon) = 
\begin{cases}
\frac{\chi_{L}^{(\pm)}(x')\chi_{R}^{(\pm)}(x) }{A\,W^{(\pm)}(x')}  & \text{for}~x > x'~; \\
\frac{\chi_{L}^{(\pm)}(x)\chi_{R}^{(\pm)}(x') }{A\,W^{(\pm)}(x')}  & \text{for}~x < x'~.
\end{cases}
\end{equation}
with the Wronskian
\[ 
W^{\pm}(x') = \chi_{L}^{\pm}(x')\frac{ d \chi_{R}^{\pm}(x')}{d x'} -
\chi_{R}^{\pm}(x')\frac{ d \chi_{L}^{\pm}(x')}{d x'}~.
\]
Inserting the result for the Green's function in Eq.~(\ref{eq:trans1st}) and using that $k_+=i\kappa$ and  $k_-=(k_+)^*$, we obtain Eqs.~(\ref{eq:transNEGF}) and (\ref{eq:tqe}) after restoring the $\bq$-dependence and taking $\tilde \eta^R=\tilde \eta^L=\tilde \eta$.

\section{Stochastic Formalism for Spin Transport in a Ferromagnet}
\label{app:stochastic}
Here, we show how to recover our analytical results from the stochastic Landau-Lifshitz-Gilbert equation, generalizing the results of Ref.~[\onlinecite{hoffman2013landau}] to the case of nonzero spin accumulation in the metallic reservoirs. 
The dynamics of the spin density unit vector $\mathbf{n}$ is governed by: 
\begin{equation}
\left(1+\alpha\mathbf{n}\times\right)\hbar\dot{\mathbf{n}}+\mathbf{n}\times\left(\mathbf{H}+\mathbf{h}\right)-A\mathbf{n}\times\nabla^{2}\mathbf{n}=0,\label{eq:bulk}
\end{equation}
where $\mathbf{H}=\Delta\hat{\mathbf{z}}$ is the effective applied magnetic
field (in units of energy) and $\mathbf{h}$ is the bulk stochastic field \cite{brown1963thermal}.
We assume a spin accumulation $\boldsymbol{\mu}'=\Delta\mu_R\mathbf{z}$ in the right normal metal, while the spin accumulation in the left lead is taken zero. The boundary condition at $x=0$ reads
\begin{eqnarray}
&&\mathbf{j}_{s}\left(x=0\right)=-\left.A\tilde s\mathbf{n}\times\partial_{x}\mathbf{n}\right|_{x=0}
\nonumber \\
&& =\left[\frac{g^{\uparrow\downarrow}}{4\pi}\left(\mathbf{n}\times\left(\mathbf{n}\times\boldsymbol{\mu}'\right)+\mathbf{n}\times\hbar\dot{\mathbf{n}}\right)+\mathbf{n}\times\mathbf{h}'_{L}\right]_{x=0}\label{eq: BCL1}
\end{eqnarray}
and at $x=d$:
\begin{eqnarray}
&& \mathbf{j}_{s}\left(x=d\right)=-\left.A\tilde s\mathbf{n}\times\partial_{x}\mathbf{n}\right|_{x=d}\nonumber \\
&& =
-\left[\frac{g^{\uparrow\downarrow}}{4\pi}\left(\mathbf{n}\times\hbar\dot{\mathbf{n}}\right)+\mathbf{n}\times\mathbf{h}'_{R}\right]_{x=d}.\label{eq: BCR1}
\end{eqnarray}
Defining $\psi\left(\mathbf{x},t\right)=n\left(\mathbf{x},t\right)\sqrt{\tilde s/2}$,
where $n\equiv n_{x}-in_{y}$, we linearize the dynamics around the
equilibrium orientation $\mathbf{n}=-\mathbf{z}$. Fourier transforming:
\[
\psi\left(x,\mathbf{q},\epsilon \right)=\int\frac{d t}{2\pi\hbar }\int\frac{d^{2}\boldsymbol{r_{\perp}}}{2\pi}e^{i\epsilon t/\hbar}e^{-i\boldsymbol{r_{\perp}}\cdot\mathbf{q}}\psi\left(x,\boldsymbol{r_{\perp}},t\right),
\]
the bulk equation of motion reads: 
\begin{equation}
A\left(\partial_{x}^{2}-\kappa^{2}\right)\psi=h\sqrt{\tilde s}~.\label{eq: bulk transformed}
\end{equation}
The bulk transformed stochastic field $h=h_{x}-ih_{y}$ obeys the
fluctuation dissipation theorem:
\begin{eqnarray}
&& \left\langle h^{*}\left(x,\mathbf{q},\epsilon \right)h\left(x',\mathbf{q}',\epsilon'\right)\right\rangle =2\left(2\pi\right)^{3}\alpha(\hbar^2/\tilde s)\epsilon
\nonumber \\
&& \times \frac{\delta\left(x-x'\right)\delta\left(\mathbf{q}-\mathbf{q}'\right)\delta\left(\epsilon-\epsilon'\right)}{{\rm tanh}\left[\epsilon/2 k_BT\right]}\, .\label{eq: bulk field}
\end{eqnarray}

The boundary conditions, Eqs. (\ref{eq: BCL1}) and (\ref{eq: BCR1}),
become respectively: 
\begin{equation}
A\partial_{x}\psi-i\frac{g^{\uparrow\downarrow}}{4\pi \tilde s}\left(\epsilon-\Delta\mu_R\right)=\frac{h_{R}}{\sqrt{2\tilde s}}\label{eq: left bc}
\end{equation}
at $x=0$  and 
\begin{equation}
A\partial_{x}\psi+i\frac{g^{\uparrow\downarrow}}{4\pi \tilde s}\epsilon\psi=\frac{h_{L}}{\sqrt{2\tilde s}}\label{eq: right bc}
\end{equation}
at $x=d$, where we have taken the coupling at both interfaces equal. Similarly, the interfacial stochastic fields obey:
\begin{eqnarray}
&&\left\langle  h_{R}^{'*}\left(\mathbf{q},\epsilon\right)h'_{R}\left(\mathbf{q}',\epsilon'\right)\right\rangle \nonumber \\
&&=\frac{2\left(2\pi\right)^{3}\alpha'd\hbar^2 \tilde s\left(\epsilon-\Delta\mu_R\right)\delta\left(\mathbf{q}-\mathbf{q}'\right)\delta\left(\epsilon-\epsilon'\right)}{{\rm tanh}\left[(\epsilon-\Delta\mu_R)/2 k_BT\right]}\label{eq: left stoch}
\end{eqnarray}
and
\begin{eqnarray}
&& \left\langle h_{L}^{'*}\left(\mathbf{q},\epsilon'\right)h'_{L}\left(\mathbf{q}',\epsilon\right)\right\rangle 
\nonumber \\
&&=\frac{2\left(2\pi\right)^{3}\alpha'd\hbar^2 \tilde s\epsilon\delta\left(\mathbf{q}-\mathbf{q}'\right)\delta\left(\epsilon-\epsilon'\right)}{{\rm tanh}\left[\epsilon/2 k_BT\right]}.\label{eq: right stoch}
\end{eqnarray}

Using Eqs. (\ref{eq: bulk transformed})-(\ref{eq: right stoch}),
one finds the current on the left side of the structure:
$j_{s}^L\equiv\mathbf{z}\cdot\mathbf{j}_{s}\left(x=0\right)$ to be of the form:
\begin{equation}
j_{s}^L=\int \frac{d \epsilon  }{2 \pi }   \, \left[N_{B}\left(\frac{\epsilon}{k_{B}T}\right)-N_{B}\left(\frac{\epsilon-\Delta\mu_{R}}{k_{B}T}\right)\right]{\rm T}(\epsilon)
\label{jcl}
\end{equation}
where ${\rm T} (\epsilon)$ is the transmission coefficient in Eq.~(\ref{eq:transNEGF}). Hence, for a clean system and in the continuum limit the results of the stochastic Landau-Lifshitz-Gilbert equation coincide with those of the NEGF formalism given by Eq.~(\ref{eq:lb_current}). 

\end{widetext}

\end{document}